\documentclass[onecolumn]{emulateapj}

\begin{document}
\title{Inflow-Outflow Solution with Stellar Winds and Conduction near Sgr A*}   
\author{Roman V. Shcherbakov\altaffilmark{1}, Frederick K. Baganoff\altaffilmark{2}}
\altaffiltext{1}{Harvard-Smithsonian Center for Astrophysics, 60 Garden Street, Cambridge, MA
02138, USA} \altaffiltext{2}{Center for Space Research, Massachusetts Institute of Technology,
Cambridge, MA 02139, USA} \email{rshcherbakov@cfa.harvard.edu}

\begin{abstract}
We propose a 2-temperature radial dynamical model of plasma flow near Sgr A* and fit the
bremsstrahlung emission to extensive quiescent X-Ray Chandra data. The model extends from several
arcseconds to black hole (BH) gravitational radius, describing the outer accretion flow together
with the infalling region. The model incorporates electron heat conduction, relativistic heat
capacity of particles and feeding by stellar winds. Stellar winds from each star are considered
separately as sources of mass, momentum and energy. Self-consistent search for the stagnation and
sonic points is performed. Most of gas is found to outflow from the region. The accretion rate is
limited to below $1\%$ of Bondi rate due to the effect of thermal conduction enhanced by entropy
production in a turbulent flow. The X-Ray brightness profile proves too steep near the BH, thus a
synchrotron self-Compton point source is inferred with luminosity $L\sim3\cdot10^{32}$~erg/s. We
fit the sub-mm emission from the inner flow, thus aiming at a single model of Sgr A* accretion
suitable at any radius.
\end{abstract}


\section{Introduction}
Our Galaxy is known to host a supermassive black hole (BH) with mass $M\approx4.5\cdot10^6 M_\odot$
at a distance $R\approx8.4$~kpc \citep{ghez}. The BH exhibits very low luminosity state probably
due to inefficient feeding and cooling. The BH is fed by stellar winds within several arcseconds
from the compact object roughly around the radius of BH gravitational influence \citep{cuadra}. The
stellar winds are expelled at large speeds. They collide, heat up to $\sim10^7$~K and emit
bremsstrahlung X-Rays, observed by Chandra \citep{baganoff}. A small fraction of mass accretes onto
the black hole is thus producing the emission in sub-mm and other wavebands. However, the inferred
accretion rate within several Schwarzschild radii is 2 orders of magnitude lower \citep{marrone}
than the inferred Bondi accretion rate \citep{bondi52} at several arcseconds. This disparity is
resolved in a present work with a point source revealed coincident with Sgr A*. A brief account of
observations is made in \S~\ref{s_obs}. The dynamical model is outlined in \S~\ref{s_model}. The
results are discussed in \S~\ref{s_res}.

\section{Observations}\label{s_obs}
We analyze $\sim1$~Ms of Chandra exposure of Sgr A* and central arcseconds \citep{muno}
significantly improving over the previously analyzed $41$~ks exposure \citep{baganoff}. As we are
interested in the quiescent emission, we bin the observations in $628$~s bins and exclude the
flaring states with counts triple the mean. We also subtract the point sources and extended bright
emission zones, thus extracting the quiescent count surface brightness profile within $5''.$ Having
the extensive data we are able to perform the subpixel spatial binning in rings $0.125''$ thickness
owing to dithering of spacecraft. The counts from four $90\deg$ ring segments centered at Sgr A*
are compared in order to test the viability of the radial model. It appears that within $2''$ the
counts do not differ significantly between ring segments, but the variation was found at $>2''.$
The point spread function (PSF) is extracted by observing the nearby binary J174540.9-290014.
\section{Stellar Winds and Dynamical Model}\label{s_model}
\begin{figure}[!ht]
\plottwo{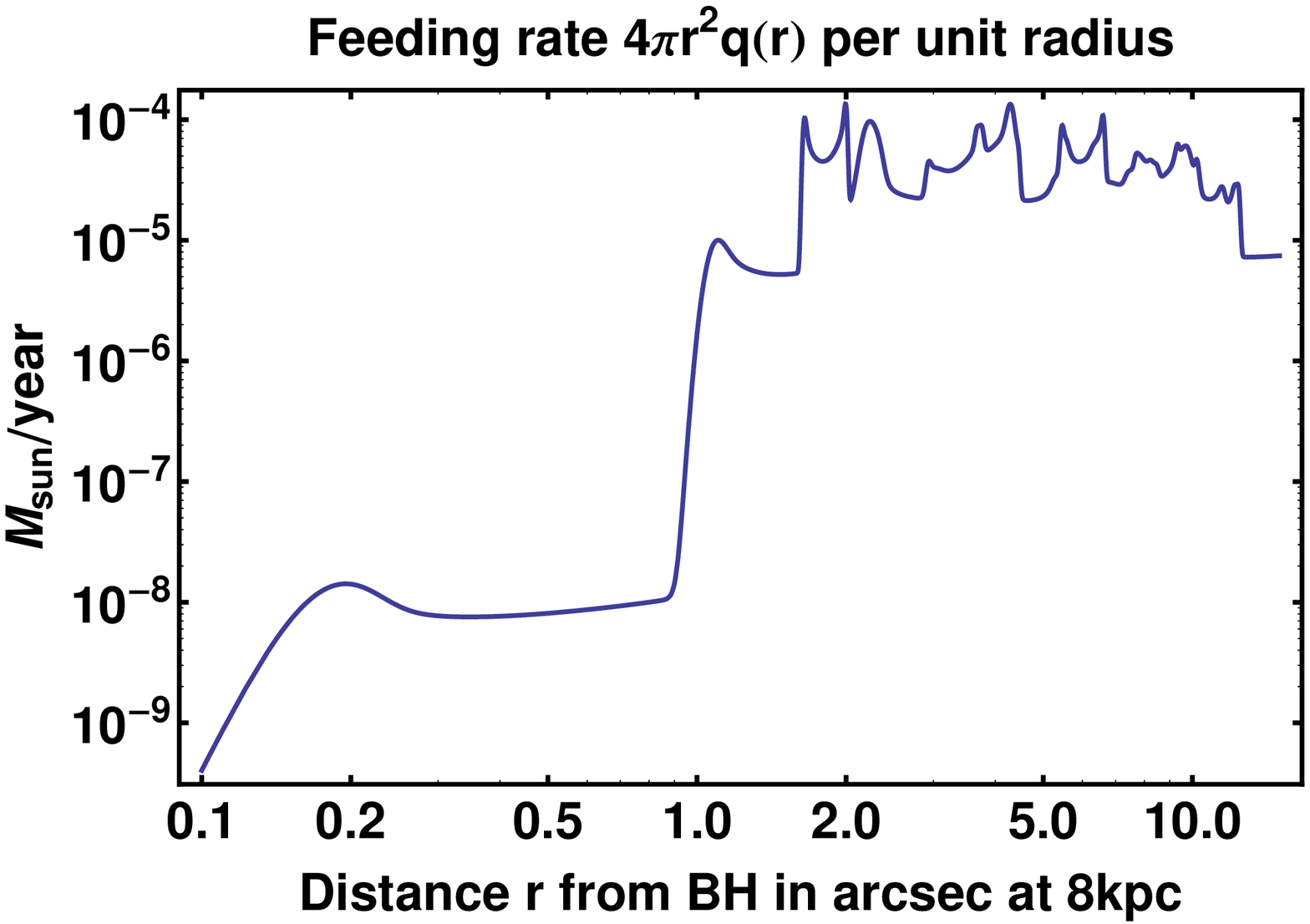}{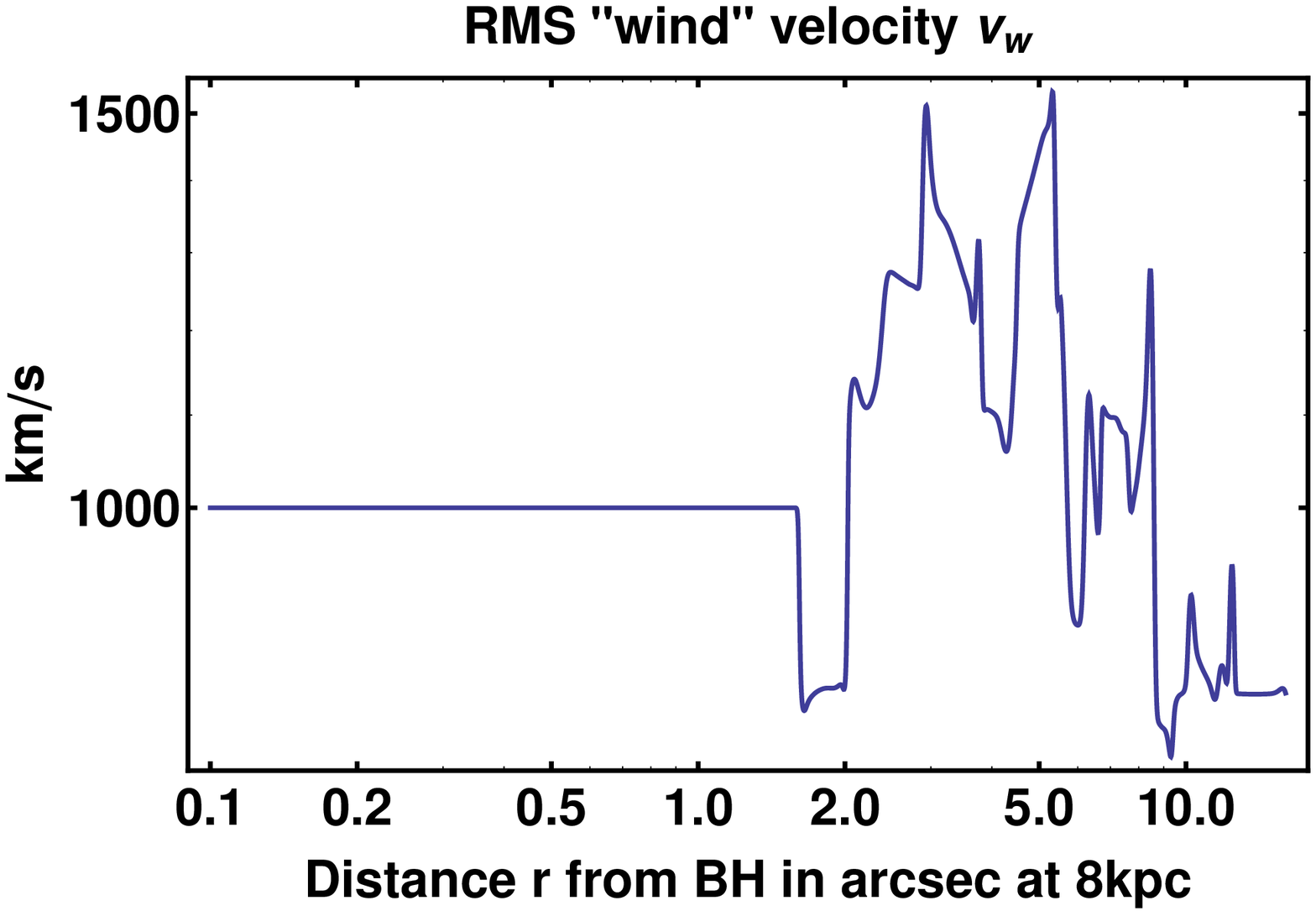}
 \caption{Mass input into the central region around the BH on the left panel. Root-mean-square wind velocity $v_w$ on the right panel.}
 \label{fig_feed}
\end{figure}

Feeding of the black hole should be a starting point of any accretion model. This approach helps to
eliminate a number of arbitrary boundary conditions. A set of $\sim30$ wind emitters is believed to
supply almost all the matter into the feeding region of Sgr A*. Following \citet{cuadra}, we
identify the important wind emitters, find the wind speeds and ejection rates. We obtain the
orbital data from \citet{paumard, martins,lu_ghez}, assuming the stars either belong to the disk or
taken to have the minimum eccentricities.  As we are constructing the radial model, the radial
feeding function $q(r)$ is produced by smoothing wind inputs over radius between the apocenter and
the pericenter for each star (see Fig.~\ref{fig_feed}). The averaged wind velocity is found as a
root-mean-square average over stars weighed with the ejection rate. We do not account for orbital
velocities of stars in energy input as feeding is dominated by only a few stars close to the BH. S2
star is included into the calculation as it may eject more matter \citep{martins_s2}, than falls
onto the BH.

The dynamical model has sources of mass, radial momentum and energy due to winds \citep{lamers}.
The main feature of the model is the electron thermal conduction proportional to the temperature
gradient with conductivity $\kappa=0.1 \sqrt{k_B T_e/m_e}n_e r$ \citep{johnson}. Here we have
included the factor of $5$ inhibition of conductivity in turbulent magnetic field \citep{medvedev}.
So computed heat flux appears to be below the saturated flux \citep{cowie}. In reality the inner
accretion flow is collisionless and the outer accretion is weakly collisional, however, the single
prescription for conductivity gives the reasonable approximation. The realistic behavior of
electrons in the inner flow is achieved by using their relativistic heat capacity in the equations
of motion, which naturally leads to the ratios $T_p/T_e$ up to 10 even in adiabatic flows. For
completeness of the theory we add the direct energy transfer into electrons and protons equivalent
to the entropy production, which happens due to viscosity in the rotating flow or due to
dissipation of turbulence \citep{shcher}. We assume that the fractions $f_e$ and $f_i$ of available
gravitational energy goes to electrons and protons. The Coulomb collisions are included for
numerical stability to balance the electron and proton temperature in the outer flow, though they
do not have the significant dynamical effect. The effect of gas composition is accounted for by
introducing the effective mass $m_{av}\approx 1.245m_H$ per electron and correspondent reduction in
ion gas pressure. These numbers are taken for the solar abundance of elements, which seems to be
reasonable for stellar winds \citep{najarro}. Paczhynski-Wiita gravitational potential is employed.

The proposed system of equations has no artificial boundary conditions, but it appears to have an
unmatched complexity. We self-consistently solve for the positions of the stagnation point, where
gas velocity is zero, and the inner isothermal sonic point \citep{quataert_wind}. The heat flux is
set to zero at the point, where $dT_e/dr=0$ in the Bondi solution near the BH. The outer boundary
is either taken to be the isothermal sonic point in the outflow or the point with slightly higher
density (for numerical stability). The relaxation technique is used for the 2-temperature system
between the inner boundary and the stagnation point, whereas only shooting works outside the
stagnation point.

\section{Results}\label{s_res}
Having produced a bunch of dynamical models, we convert the temperature and density radial profiles
into the surface brightness profile. We take the up-to-date bremsstrahlung emissivities (see
\citep{gould} and errata) and account for emission by heavy elements, excluding iron.
\begin{figure}[!ht]
\plottwo{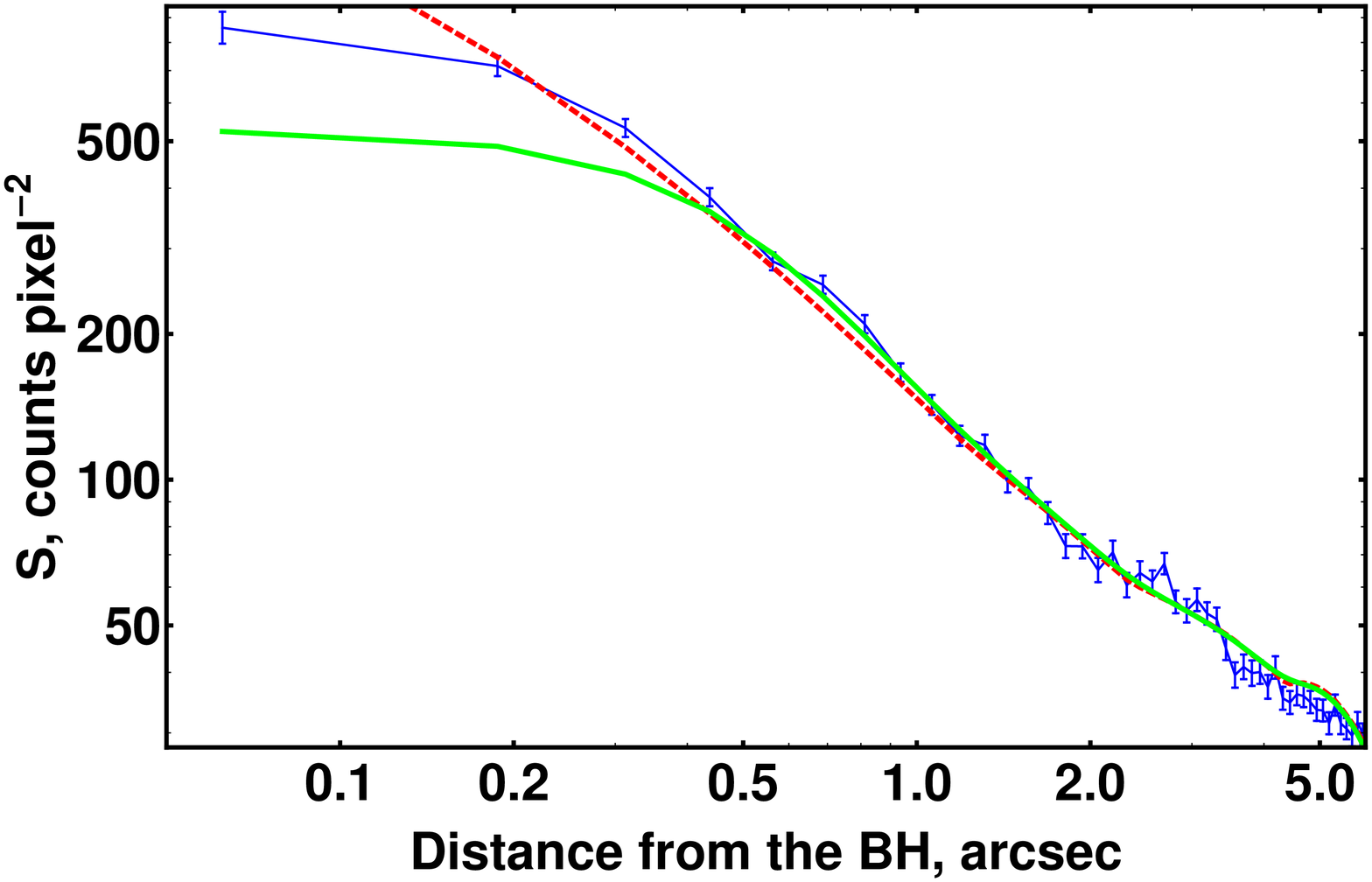}{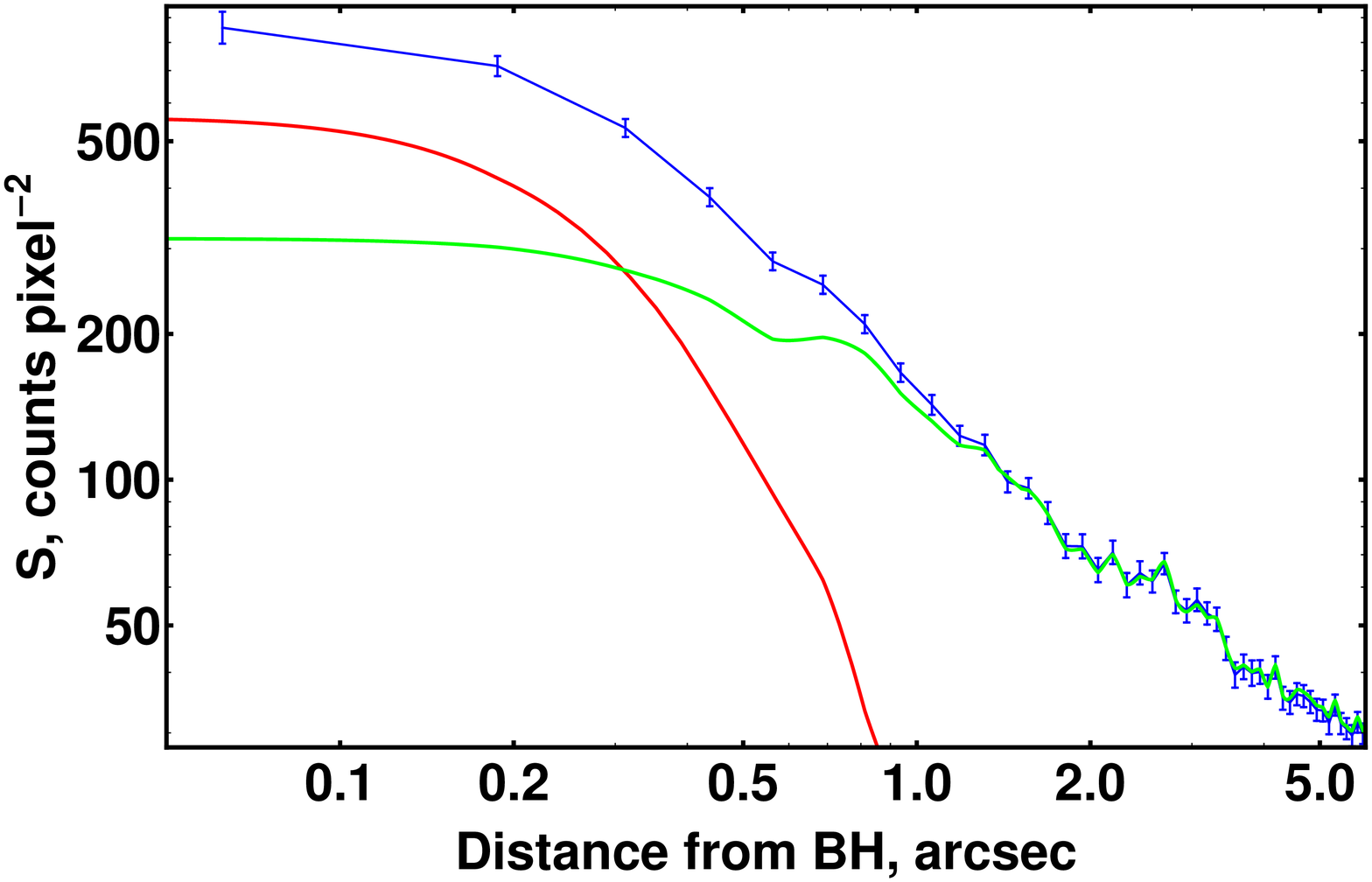} \caption{Plots of surface brightness S in observed counts per
pixel squared (1 pixel=$0.5''$). Blue curve with error bars shows the observations. Extended
emission model is on the left panel: the green curve shows brightness smoothed by PSF. The model of
a point source with $L=3\cdot10^{32}$~erg/s (red) and the residual (green) are on the right panel.}
\label{results}
\end{figure}We calculate
the spectrum along each ray and convert it to counts by applying the solar metallicity interstellar
absorption \citep{morrison} with hydrogen column $N_H=10^{23}$cm${}^{-2}$ and convolving with the
response of Chandra. Then we apply the PSF blurring and compare with the observed surface
brightness profile. The outer surface brightness profile can be reasonably fitted by a model with
$f_e=0.15,$ $f_i=0.05,$ which also leads to the ratio $T_p/T_e\approx15$ near the BH. However, we
find that the inner part of surface brightness curve is too steep for any extended emission. There
should be a point source in the center accounting for $\onethird$ to $\twothirds$ of central
surface brightness and having the unabsorbed luminosity about $L=3\cdot10^{32}$~erg/s of
monoenergetic $4$~keV photons. About the same luminosity is expected from the synchrotron
self-Compton (SSC) process near the BH. The search for the best model with the point source
continues. We want the best model to reproduce the observed Faraday rotation measure
$RM\approx50{\rm cm}^{-2}$\citep{marrone} and optically thick flux $F_R=1.73$~Jy at $86$~GHz
\citep{krichbaum}. An order of magnitude consistency is achieved on the way. The accretion rate
appears to be self-consistently limited to $<1\%$ of Bondi value, thus the connection between the
inner accretion flow and the outer accretion flow is established. The future versions will include
the angular momentum and use X-Ray spectral information.

 \acknowledgements The
author is grateful to Mikhail Medvedev and Ramesh Narayan for fruitful discussions, Daniel Wang and
Feng Yuan for comments.

\end{document}